\documentclass[letterpaper, 10 pt, conference]{ieeeconf}  

\IEEEoverridecommandlockouts                              

\usepackage{cite}
\usepackage{amsmath,amssymb,amsfonts}
\usepackage{graphicx} 
\usepackage{amsmath}
\usepackage{amssymb}  
\usepackage{pifont}
\newcommand{\cmark}{\ding{51}}%
\newcommand{\xmark}{\ding{55}}%
\usepackage{mathrsfs}
\usepackage{multicol}
\usepackage{xcolor}
\usepackage{float}
\usepackage{booktabs}
\usepackage{subcaption}
\usepackage{mathtools}
\usepackage{soul}
\usepackage[table]{xcolor}
\usepackage{cancel}
\usepackage{bm}
\usepackage[T1]{fontenc}
\newtheorem{rem}{Remark}

\newtheorem{defn}{Definition}
\newtheorem{prop}{Proposition}

\newtheorem{theorem}{Theorem}

\usepackage{tikz}
\usepackage{tcolorbox}
\usepackage{dsfont}
\usepackage{algorithm}
\usepackage{algpseudocode}
\usepackage{hyperref} 
\newcommand{\mc}{\mathcal}
\let\labelindent\relax
\usepackage[shortlabels]{enumitem}

\usepackage{wrapfig}
\usepackage{tcolorbox}
\usepackage{cleveref}

\usepackage{dsfont}

\newtcolorbox[auto counter,number within=section,crefname={box}{boxes}]{pabox}[2][]{%
title=Box 1 $\mid$ Timeline,
colback=gray!20,
colframe=gray!50,
fonttitle=\bfseries,
}

\makeatletter
\@ifundefined{proof}{%
\newenvironment{proof}{\paragraph{Proof:}}{\hfill$\square$}%
}{}
\makeatother

\DeclareMathAlphabet\mathbfcal{OMS}{cmsy}{b}{n}

\definecolor{darkspringgreen}{rgb}{0.09, 0.45, 0.27}

\def\BibTeX{{\rm B\kern-.05em{\sc i\kern-.025em b}\kern-.08em
    T\kern-.1667em\lower.7ex\hbox{E}\kern-.125emX}}

\def\1{\mathds{1}}
\def\C{\mathbb{C}}
\def\R{\mathbb{R}}

\newcounter{myblockcounter}
\renewcommand{\themyblockcounter}{BLK-\arabic{myblockcounter}}

\newtcolorbox[use counter=myblockcounter]{mybox}[2]{%
  colback=gray!20, 
  colframe=gray!50, 
  sharp corners=south, 
  title=Block \themyblockcounter: #1, 
  label=#2
}

\title{\LARGE \bf
 L2RU: a Structured State Space Model with prescribed $  \mc L_2$-bound 
}

\author{Leonardo Massai, Muhammad Zakwan and Giancarlo Ferrari-Trecate
\thanks{This research has been supported by the Swiss National Science Foundation under the NCCR Automation (grant agreement 51NF40\_180545).}
\thanks{The authors are with the Institute of Mechanical Engineering, Ecole Polytechnique Fédérale de Lausanne (EPFL), CH-1015 Lausanne, Switzerland. (email: \tt\small {\{l.massai,  giancarlo.ferraritrecate\}@epfl.ch) } }%
}

\begin{document}

\maketitle
\thispagestyle{empty}
\pagestyle{empty}

\begin{abstract}
Structured state-space models (SSMs) have recently emerged as a powerful architecture at the intersection of machine learning and control, featuring layers composed of discrete-time linear time-invariant (LTI) systems followed by pointwise nonlinearities. These models combine the expressiveness of deep neural networks with the interpretability and inductive bias of dynamical systems, offering remarkable computational efficiency and strong performance on long-sequence tasks. However, their adoption in applications such as system identification and optimal control remains limited by the difficulty of ensuring stability and robustness in a principled and tractable manner.
We introduce L2RU, a novel class of SSMs endowed with a prescribed $\mathcal{L}_2$-bound, guaranteeing input–output stability and robustness for all parameter values. The L2RU architecture is derived from a family of free parametrizations of LTI systems satisfying an $\mathcal{L}_2$ constraint, which allow unconstrained optimization via standard gradient-based methods while maintaining rigorous stability guarantees. Specifically, we develop two complementary parametrizations: a non-conservative formulation that provides a complete characterization of all square LTI systems with a given $\mathcal{L}_2$-bound, and a conservative formulation that extends the approach to general (possibly non-square) systems while enabling greater computational efficiency through a structured representation of the system matrices.
Both parametrizations admit efficient initialization schemes that facilitate the training of long-memory models. We demonstrate the effectiveness of the proposed framework on a nonlinear system identification benchmark, where L2RU achieves superior performance and training stability compared to existing SSM architectures, underscoring its potential as a principled and robust building block for learning and control.
\end{abstract}


\section{Introduction}

In recent years, a surge of research interest has been witnessed in deep-learning foundation models for control. A wide range of increasingly sophisticated architectures, from Recurrent Neural Networks (RNNs) \cite{bonassi_recurrent_2022, lanzetti_recurrent_2019, andersson_deep_2019} to Transformers \cite{vaswani_attention_2023, sun_efficient_2022}, have been proposed for nonlinear system identification and optimal control, where they serve as parametrizations for highly nonlinear controllers. A significant portion of this research has focused on developing parametrizations that enforce specific stability properties, which are often critical in control applications. For example, in system identification, ensuring a priori that learned dynamical models possess guaranteed stability and robustness is essential whenever the system generating the data exhibits these properties. Similarly, in optimal control the search space is typically constrained to stabilizing controllers.
Several parameterization techniques have been explored to achieve these stability guarantees. In \cite{pauli_training_2022, wang_direct_2023} Neural Networks (NNs) and convolutional NNs with prescribed tight Lipschitz-bounds are studied, while \cite{revay_recurrent_2024} introduces Recurrent Equilibrium Networks (RENs), a class of RNNs with guaranteed dissipativity properties, ensuring finite $\mc L_2$-gain and contractivity.

Another class of models that has recently gained significant attention in machine learning and control are the Structured State-space Models (SSMs). The interest in SSMs, an architecture composed of multiple layers of LTI discrete-time systems followed by nonlinear functions, was ignited by the work in \cite{gu_efficiently_2022}, where the S4 architecture was introduced. SSMs are computationally efficient due to algorithms such as Parallel Scan \cite{blelloch_prefix_2004} and have been shown to offer performance on long-context tasks comparable to state-of-the-art Transformers \cite{gu_mamba:_2024}. Since then, numerous variants have been proposed, leveraging different discretization schemes, nonlinearities and parametrizations for the underlying LTI systems \cite{smith_simplified_2023} (see \cite{alonso_state_2024} for a comprehensive survey). Notably, in \cite{orvieto_resurrecting_2023} the authors investigated a direct discrete-time parameterization of LTI subsystems, leading to the Linear Recurrent Unit (LRU) architecture.
Furthermore, due to the relatively simple structure of the recurrent component (the LTI system), SSMs are more interpretable and amenable to formal analysis compared to architectures like Transformers, especially when viewed through the lens of control theory. Notably, all SSM architectures mentioned above enforce certain forms of input-to-state stability by directly parametrizing stable LTI systems, as discussed in \cite{bonassi_structured_2024}.

\subsubsection*{Contributions}
Motivated by the desire to combine the efficiency and learning power of SSMs with the need for strong stability guarantees, we introduce a novel parametrization of SSMs that guarantees input/output stability and robustness, certified by a prescribed $\mc L_2$-bound, which quantifies the worst-case amplification of the output for any finite input. We refer to our architecture as L2RU, highlighting both its connection to the concept of $\mc L_2$-gain and its similarities to the LRU introduced in \cite{orvieto_resurrecting_2023}. The ability to enforce an $\mc L_2$-bound a priori is crucial for applications requiring strict robustness guarantees, such as system identification and optimal control with stability constraints \cite{furieri_learning_2024}. This holds in both centralized and distributed settings \cite{furieri_distributed_2022}, where models with prescribed $\mc L_2$-bounds enable the construction of networked stable systems \cite{massai_unconstrained_2024, saccani_optimal_2024}. Furthermore, in learning tasks, a certifiable and quantifiable robustness property enhances the resilience of the model against adversarial attacks \cite{zakwan_robust_2023}.
Importantly, the proposed parametrization is free in the sense that the prescribed $\mc L_2$-bound is guaranteed for all parameter values, eliminating the need for complex constraints. This allows L2RUs to be optimized via unconstrained optimization techniques, such as stochastic gradient descent and off-the-shelf automatic differentiation tools.

Our contributions are threefold.
First, we develop two free parametrizations of discrete-time LTI systems that satisfy a prescribed $\mathcal{L}_2$-bound: a non-conservative parametrization that provides a complete characterization of square systems, and a conservative one that extends the formulation to general (possibly non-square) systems while also enjoying superior computational efficiency.
Second, we leverage these results to construct the proposed L2RU architecture, an $\mathcal{L}_2$-bounded SSM layer that can be trained through unconstrained optimization while guaranteeing input–output stability and robustness certified by a prescribed $\mc L_2$-bound.
Third, we introduce an initialization strategy specifically designed to enhance memory retention and improve performance when processing long input sequences.
Finally, we validate the proposed approach through a system identification benchmark, demonstrating its effectiveness and robustness compared to existing SSM architectures.

\section*{Notation}
Throughout the paper, vectors are denoted with lowercase, matrices with uppercase, and sets with calligraphic letters. Sequences of vectors are denoted with bold lowercase and the set of all sequences $\mathbf{v}=(v_0,v_1,v_2,\dots)$, where $v_t\in\mathbb{R}^n$ for all $t\in \mathbb{N}$, is denoted with $\mc L^n$. Moreover, $\mathbf{v}$ belongs to the set of square-integrable sequences $\mc L_2^n \subset \mc L^n$ if $\| \mathbf{v} \|_2=\left( \sum_{t=0}^{\infty} \|v_t\|_2^2 \right)^{\frac{1}{2}}<\infty$. The set of $n\times n$ orthogonal matrices is denoted with $\mc O(n)$ and the group of special orthogonal matrices with positive determinant with $\mc S \mc O(n)$. The expression $A \succ0$ ($A \succeq0$) defines a positive (semi) definite matrix $A$. The spectrum (set of eigenvalues) of $A$ is denoted with $\lambda(A)$, its partition into conformal blocks with $A=\operatorname{Blk}(A_{11}, A_{12}, A_{21}, A_{22})$ and, for $A \succ 0$, its Cholesky decomposition with $A=L_AL_A^{\top}$. The identity matrix is indicated with $I$, regardless of its dimension. We indicate the logistic function with $\sigma(x):=\frac{1}{1+e^{-x}}$. Finally, whenever convenient, we use the natural isomorphisms $\mathbb{R}^n \times \mathbb{R}^m \cong \mathbb{R}^{n+m}$ and $\mathbb{R}^{n \times m} \cong \mathbb{R}^{nm}$, identifying elements via concatenation and column-stacking, respectively.


\section{Preliminaries} \label{sec:preliminaries}

A foundation model can be seen as a map $f_{\theta}:\mc L^{n_u} \mapsto \mc L^{n_y}$, with $n_u,n_y\in\mathbb N$, depending on a parameter $\theta\in\R^n$ and differentiable with respect to it. Given an input sequence $\mathbf u\in\mc L^{n_u}$, the output is $\mathbf y=f_\theta(\mathbf u)$. This map can be static, as in a memory-less Multi-Layer Perceptron (MLP), or dynamical, as in recurrent architectures. The specific architecture defining $f_\theta$ is a design choice that depends on the problem at hand.

In control-oriented learning tasks, the parameter set is often chosen so that $f_\theta$ satisfies a desired stability or robustness property for all admissible values of $\theta$. The property considered in this paper is a prescribed input-output $\mc L_2$-bound.

\begin{defn}($\mc L_2$-\emph{gain}) \label{def:l2}
   Let $f: \mc L_2^{n_u} \mapsto \mc L_2^{n_y}$ be a map between square-integrable sequences. The map $f$ has finite $\mc L_2$-gain if there exists $\gamma>0$ such that, for any $\mathbf u\in\mc L_2^{n_u}$,
    \begin{equation}\label{eq:l2bound}
        \| f(\mathbf u)\|_2 \le \gamma \|\mathbf u\|_2 .
    \end{equation}
  Any positive $\gamma$ for which~\eqref{eq:l2bound} holds is called an $\mc L_2$-bound of $f$. The $\mc L_2$-gain of $f$ is the infimum among all such bounds.
\end{defn}

When $f$ is described by a dynamical system, the input-output estimate may contain an additional term depending on the initial state~\cite{caverly_lmi_2024}. In this paper, the recurrent linear systems are initialized at zero when defining their input-output gain, so this term vanishes.

For LTI systems, Definition~\ref{def:l2} admits the following standard characterization.

\begin{prop}\label{prop:boundedlemma} (\emph{DT Real Bounded Lemma}~\cite{caverly_lmi_2024})
Let $g_{(A,B,C,D)} : \mc L^{n_u} \mapsto \mc L^{n_y}$ be a DT LTI system described in state space. The system $g$ has $\mc L_2$-bound $\gamma$ if and only if there exists $P\succ0$ such that
\begin{equation}\label{eq:realbounded}
\left[\begin{array}{cc}
A^{\top} P A-P+C^{\top} C & A^{\top} P B+C^{\top} D \\
B^{\top} P A+D^{\top} C & B^{\top} P B+D^{\top} D-\gamma^2 I
\end{array}\right] \prec 0 .
\end{equation}
Equivalently, the same condition can be written as
\begin{equation}
    \label{eq:realbounded2}
\left[\begin{array}{cccc}
P & P A  & P B & 0  \\
A^\top P & P  & 0 & C^\top \\
B^\top P & 0  & \gamma I & D^\top \\
0 & C & D & \gamma I
\end{array}\right] \succ 0 .
\end{equation}
Moreover, the $\mc L_2$-gain of $g$ is obtained by minimizing $\gamma$ subject to either of the above LMIs.
\end{prop}

The goal of using such certificates in learning is to avoid constrained optimization over implicitly defined sets of matrices. This motivates the following notion.

\begin{defn}(\emph{Parametrization})\label{def:param}
Let $\mc B\subseteq\mathbb R^m$ be a nonempty set. A \emph{parametrization} of $\mc B$ is a differentiable map
$
\psi:\mathcal A\subseteq\mathbb R^n\mapsto \mc B.
$
We say that $\psi$ is \emph{free} if there exists a set $\mathcal N\subset\mathbb R^n$ with zero Lebesgue measure such that $\mathcal A=\mathbb R^n\setminus\mathcal N$, i.e., $\psi$ is defined and differentiable almost everywhere on $\mathbb R^n$.
We say that $\psi$ is \emph{complete} if there exists a set $\mathcal M\subset\mc B$ of zero Lebesgue measure such that $\psi(\mathcal A)=\mc B\setminus\mathcal M$.
\end{defn}

We can think of $\mc B$ as the set of parameters ensuring that the model $f_\theta$ satisfies a given property. A free parametrization guarantees this property for almost all values of the unconstrained parameter. In a learning task with input-output pairs $(\tilde{\mathbf u},\tilde{\mathbf y})$ and loss function $l$, it transforms the constrained problem
$\min_{\theta\in\mc B} l(\tilde{\mathbf y},f_\theta(\tilde{\mathbf u}))$
into the unconstrained one
$\min_\omega l(\tilde{\mathbf y},f_{\psi(\omega)}(\tilde{\mathbf u}))$.
A complete parametrization additionally avoids conservatism, up to the null sets allowed in Definition~\ref{def:param}.

\begin{rem}
In Definition~\ref{def:param} we allow neglecting null sets for practical reasons in learning. The parameter is typically initialized randomly and then adjusted by gradient-based optimization. Since the parametrization is defined almost everywhere, the iterates avoid the singular sets with probability one under random initialization and arbitrarily small random perturbations. Likewise, allowing the image to miss a null set is harmless for approximation purposes, since the missing points can be approached arbitrarily closely by admissible parameter values.
\end{rem}

\section{The L2RU model} \label{secII}

\begin{figure}[t]
    \centering
    \includegraphics[scale=0.8]{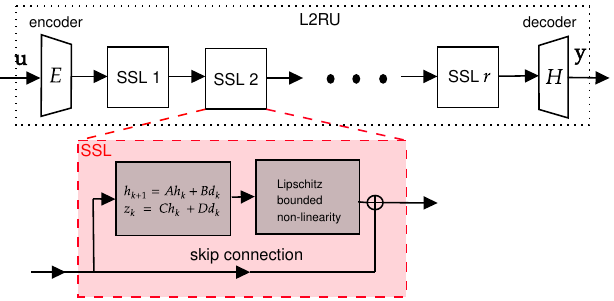}
    \caption{L2RU architecture presented in this paper. The model consists of a series of state-space layers, each comprised of $\mc L_2$-bounded DT LTI systems and Lipschitz-bounded nonlinearities. The input/output is pre- and post-processed by linear transformations.}
    \label{fig:ssml2}
\end{figure}

We now specialize the previous discussion to Structured State-space Models (SSMs)~\cite{gu_efficiently_2022}. While SSMs can take various forms, they are fundamentally characterized by layers made of discrete-time LTI systems followed by static nonlinear functions. We introduce the L2RU architecture shown in Fig.~\ref{fig:ssml2}. It resembles the Linear Recurrent Unit~\cite{orvieto_resurrecting_2023}, but differs in the way the LTI recurrent core is parametrized.

\subsection{State-space scaffold}

An L2RU is defined by the following components.
\begin{itemize}
    \item \emph{State-space layer (SSL).} Each layer is a DT LTI system followed by a static nonlinearity. The LTI system is described by
    \begin{align} \label{eq:sysl}
    g_{(A,B,C,D)}:\quad
    \begin{cases}
        h_{k+1} = A h_k+B d_k, \qquad h_0=0,\\
        z_k = C h_k+D d_k,
    \end{cases}
    \end{align}
where $h\in\R^{n_h}$ is the state, $d\in\R^{n_d}$ is the input, $z\in\R^{n_z}$ is the output, and $A,B,C,D$ have compatible dimensions. The system $g$ can be seen as a map $g:\mc L^{n_d}\mapsto\mc L^{n_z}$. A common requirement is that $A$ must be Schur, ensuring internal stability. In SSMs, this LTI system is the recurrent core that gives the model its memory.

The output of $g$ is fed to a static nonlinear function
    \begin{align} \label{eq:mu}
        \mu_{\xi,\zeta}: \R^{n_z} \mapsto \R^{n_d},
    \end{align}
where $\xi$ denotes trainable parameters and $\zeta>0$ is a prescribed Euclidean Lipschitz bound. We assume $\mu_{\xi,\zeta}(0)=0$ and
$
\|\mu_{\xi,\zeta}(a)-\mu_{\xi,\zeta}(b)\|_2\le \zeta\|a-b\|_2
$
for all $a,b\in\R^{n_z}$. Hence, when the function is applied pointwise to a sequence, $\|\mu_{\xi,\zeta}(\mathbf v)\|_2\le \zeta\|\mathbf v\|_2$. The family $\mu_{\xi,\zeta}$ can range from simple pointwise nonlinearities to Lipschitz-bounded MLPs, such as the one proposed in~\cite{wang_direct_2023}. The output of the nonlinear branch is then added to the layer input through a feed-through connection. The block can be stacked to form deep architectures.

    \item \emph{Encoder/decoder.} The input and output of the stacked SSLs are pre- and post-processed by linear maps $E\in\R^{n_d\times n_u}$ and $H\in\R^{n_y\times n_d}$. These maps act pointwise on the input and output sequences.
\end{itemize}

For a fixed number of layers $r\in\mathbb N$, the L2RU defines a dynamical map $f_\theta:\mc L^{n_u}\mapsto\mc L^{n_y}$ with parameters
\begin{align}\label{eq:setP}
    \theta \in \mc P
    &=\left\{\{\underbrace{A_i,B_i,C_i,D_i}_{g_i},\underbrace{\xi_i}_{\mu_i}\}_{1\le i\le r},E,H\right\} \notag\\
    &\subseteq \R^{r(n_h^2+n_h n_d+n_z n_h+n_z n_d+m)+n_d n_u+n_y n_d} .
\end{align}
The action of $f_\theta$ on an input sequence $\mathbf u$ can be explicitely written as
\begin{align}\label{eq:l2ruseries}
    &\text{encoder:} & \mathbf y_0 &= E\mathbf u, \notag\\
    &\text{SSLs:} & \mathbf y_i &= \mu_i\!\left(g_i(\mathbf y_{i-1})\right)+\mathbf y_{i-1},\quad 1\le i\le r,\\
    &\text{decoder:} & \mathbf y &= H\mathbf y_r . \notag
\end{align}

\subsection{From LTI certificates to an L2RU certificate}

Given a desired finite bound $\hat\gamma>0$, our goal is to parametrize the set
\begin{align}  \label{eq:omega}
    \Theta_{\hat\gamma}
    =\left\{\theta\in\mc P\;\middle|\; f_\theta\text{ has }\mc L_2\text{-bound }\hat\gamma\right\}.
\end{align}
Here, as throughout the paper, saying that a model has $\mc L_2$-bound $\hat\gamma$ means that its induced gain is not larger than $\hat\gamma$.

The construction has two steps. First, we need free parametrizations of the LTI cores with a prescribed bound. For a fixed $\gamma>0$, define
\begin{align}
    \mc G_\gamma
    =\left\{A,B,C,D\;\middle|\;\exists P\succ0\text{ satisfying }\eqref{eq:realbounded}\text{ or }\eqref{eq:realbounded2}\right\} .
    \label{eq:setG}
\end{align}
It is convenient to work with the lifted set
\begin{align}
    \mc R_\gamma
    =\left\{A,B,C,D,P\;\middle|\;P\succ0\text{ and }\eqref{eq:realbounded}\text{ or }\eqref{eq:realbounded2}\text{ holds}\right\} .
      \label{eq:setR}
\end{align}
The two sets are related by the projection
\begin{align}
 \pi(A,B,C,D,P)=(A,B,C,D). \label{eq:proj}
\end{align}
Thus, a free parametrization of $\mc R_\gamma$ immediately gives a free parametrization of $\mc G_\gamma$. Indeed, by definition, $\mc G_{\gamma} = \pi(\mc R_{\gamma})$, so the map  
$\pi \circ \psi: \mc P \mapsto \mc G_{\gamma}$  
is a free and complete parametrization of $\mc G_{\gamma}$. The reason is that $\pi$ is surjective, and hence $\pi \circ \psi$ inherits the completeness property from $\psi$. Moreover, parametrizing $\mc G_{\gamma}$ via $\mc R_{\gamma}$ yields the additional advantage of providing a stability certificate $P$ for each system defined by $(A,B,C,D)$.  

Second, once each LTI core has a known bound and each nonlinearity has a known Lipschitz constant, the complete SSM bound is obtained by composing the layer-wise estimates. The next subsections present two alternatives for the first step, and then the full L2RU scaffold.

\subsection{The complete square core $\psi_\gamma$}

Here, we present a parametrization, originally proposed in \cite{massai_free_2025}, of a specific subclass of the L2RU architecture, namely, that characterized by square LTI systems, i.e., systems for which $n_h = n_d = n_z = n$. To highlight that we are dealing with this particular case, we will denote the sets \eqref{eq:setG} and \eqref{eq:setR} by $\mc G_{\gamma}^n$ and $\mc R_{\gamma}^n$ respectively.  The square assumption is mainly technical: it enables a free and complete parametrization of the bounded-real set. It does not constrain the external input and output dimensions of the full L2RU, because $n_u$ and $n_y$ are handled by the encoder and decoder.

\begin{mybox}{Parametrization of square DT LTI systems with prescribed $\mc L_2$-bound $\gamma$}{box}
Given $\gamma>0$ and the set of free parameters
\begin{align*}
\resizebox{.97\hsize}{!}{$
 \mc F_\psi=\left\{(\alpha,\varepsilon),X_{11},X_{21},X_{22},\tilde C,\tilde D,S\right\}\simeq\R^{\frac12(9n^2+n)+2},
$}
\end{align*}
define $\psi_\gamma:\omega\in\mc F_\psi\mapsto(A,B,C,D,P)\in\R^{5n^2}$ by
\begin{align}
& A = L_{-(R-H_{11})}^{-\top}QL_{-R}^{\top}, \label{eq:pari} \\
& B = A H_{12}^{-\top}V^{\top}, \\
& C = \tilde C, \\
& D = \tilde D\sqrt{\beta}, \\
& P = -A^{-\top}H_{12}B^{-1},
\end{align}
where
\begin{align}
  &Q = \left(I-S+S^{\top}\right)\left(I+S-S^{\top}\right)^{-1}, \label{eq:orth} \\
  &Z = X_{21}X_{21}^{\top}+X_{22}X_{22}^{\top}+\tilde D\tilde D^{\top}+e^{\varepsilon}I, \label{eq:zeta} \\
  &\beta = \frac{\gamma^2\sigma(\alpha)}{\|Z\|_2}, \label{eq:beta}\\
  &H_{11}=X_{11}X_{11}^{\top}+\tilde C^{\top}\tilde C+\beta e^{\varepsilon}I, \\
  &H_{12}=\sqrt{\beta}\left(X_{11}X_{21}^{\top}+\tilde C^{\top}\tilde D\right), \\
  &V=\beta Z-\gamma^2 I,\qquad R=H_{12}V^{-\top}H_{12}^{\top}. \label{eq:pare}
\end{align}
\end{mybox}

\begin{theorem}\label{thm:param}
The map $\psi_\gamma$ defined in Block~\ref{box} is a free and complete parametrization of $\mc R_\gamma^n$ and yields one for $\mc G_\gamma^n$ through the projection~\eqref{eq:proj}.
\end{theorem}

The proof of Theorem~\ref{thm:param} is provided in~\cite{massai_free_2025}. The relevant consequence for this paper is that $\psi_\gamma$ covers almost all square LTI systems satisfying the prescribed $\mc L_2$-bound, including systems whose actual gain is arbitrarily close to $\gamma$. This makes $\psi_\gamma$ the expressive, non-conservative core of L2RU.

\begin{rem} \label{rem:complexity}
The parametrization $\psi_\gamma$ requires computing the spectral norm of an $n\times n$ symmetric matrix; see~\eqref{eq:beta}. This has complexity $O(n^3)$ and may become expensive for large $n$. Approximation methods such as power iteration can be used in practice, while upper bounds based on the Frobenius norm or Gershgorin disks are also possible. These approximations preserve the prescribed bound but generally sacrifice the completeness of the exact construction.
\end{rem}

\begin{rem} \label{rem:bound}
The prescribed bound can also be made trainable by setting $\gamma=|\tilde\gamma|$, or by using any smooth positive reparametrization of $\tilde\gamma\in\R$. This gives a free parametrization of square $\mc L_2$-bounded DT LTI systems without fixing the bound in advance. When an explicit certificate value is not needed, the complex-diagonal parametrization used in LRU-type models may be preferable computationally; the point of $\psi_\gamma$ is precisely to keep the certificate explicit and, in the square case, non-conservative.
\end{rem}

 \subsubsection*{Initialization and computational efficiency}
It is well-known in the SSMs-related literature that proper initialization of the linear system's parameters is crucial for effective learning with SSMs, particularly when dealing with long input sequences. Various initialization strategies have been proposed depending on the SSM architecture employed \cite{gu_hippo:_2020, gu_parameterization_2022, orvieto_resurrecting_2023}. Ideally, the matrix $A$ should have eigenvalues within the stability region and close to its boundary, ensuring   $|\lambda|<1, |\lambda| \approx 1 \: \forall \: \lambda \in \lambda(A)$. The intuition behind this approach is that the state evolution depends on past inputs via the powers of the matrix 
$A$, which can decay rapidly if the eigenvalues of $A$ are far from the stability region's boundary.  By ensuring $|\lambda_i| \approx 1$ at initialization, the system retains long-range memory and prevents the signal from past inputs from quickly dying out. 

For $\psi_\gamma$, the following initialization gives direct control of the initial pole modulus.

\begin{prop} \label{prop:init}
Consider
\begin{align}
\resizebox{.90\hsize}{!}{$
\omega_0=\left(\alpha,\varepsilon,X_{11}=I,X_{21}=I,X_{22}=I,\tilde C=I,\tilde D=I,S\right),
$}
\end{align}
and let $A_0$ be the matrix $A$ obtained through $\psi_\gamma(\omega_0)$ in the limit $\varepsilon\to-\infty$. Then
\begin{align}
    A_0=\sqrt{\frac{2\sigma(\alpha)}{3-\sigma(\alpha)}}\,Q,
\end{align}
and therefore
\begin{align}
    |\lambda|=\sqrt{\frac{2\sigma(\alpha)}{3-\sigma(\alpha)}}
    \qquad \forall\lambda\in\lambda(A_0),
\end{align}
where $Q$ is the orthogonal matrix in~\eqref{eq:orth}.
\end{prop}

Thanks to Proposition \ref{prop:init}, proved in Appendix \cite{massai_free_2025},  we can directly control the modulus of the eigenvalues of \( A \) at initialization. By setting a sufficiently large negative value for \( \varepsilon \) and adjusting the parameter \( \alpha \) such that \( \sigma(\alpha) \approx 1 \), we can push the eigenvalues arbitrarily close to the boundary of the unit circle. Notably, the phase of the eigenvalues is controlled by the orthogonal matrix \( Q \), and this can be adjusted by appropriately choosing the free matrix \( S \).
As we will show in the example, this initialization enhances accuracy in learning tasks by promoting long-range memory.

Finally, note that the parametrization of the matrix $A$ under $\psi_{\gamma}$, as defined in~\eqref{eq:pari}, yields a dense real matrix. The absence of a diagonal or block-diagonal structure prevents the use of efficient algorithms such as parallel scan for computing the time evolution of the associated LTI system, requiring the standard recursive simulation instead.

\subsection{The structured general core $\kappa_\gamma$}

We next introduce an alternative parametrization that applies to general, possibly non-square, LTI systems. It preserves the complex-diagonal structure of the state matrix used in LRU-type models~\cite{orvieto_resurrecting_2023}. Unlike $\psi_\gamma$, this construction is based on sufficient bounded-real conditions, and is therefore free but not complete.

\begin{mybox}{Parametrization of DT LTI systems with prescribed $\mc L_2$-bound $\gamma$}{box3}
Fix $\varepsilon_\kappa>0$. Given $\gamma>0$ and the free parameters
\begin{align*}
\resizebox{.97\hsize}{!}{$
\mc F_\kappa=
\left\{\{\mu_j,\theta_j\}_{j=1}^{n_h},\tilde D,\bar Y\;\middle|\;\mu_j,\theta_j\in\R,\tilde D\in\C^{n_z\times n_d},\bar Y\in\C^{2n_h\times(n_d+n_z)}\right\},
$}
\end{align*}
define $\kappa_\gamma:\omega\in\mc F_\kappa\mapsto(A,B,C,D)$ as follows. First set
\begin{align}
&\lambda_j=\exp\!\left(-\exp(\mu_j)+\mathrm i\exp(\theta_j)\right), \label{eq:A1}\\
&A=\operatorname{diag}(\lambda_1,\ldots,\lambda_{n_h}), \label{eq:A2}\\
&P=A^*A+\varepsilon_\kappa I_{n_h},\\
&D=\frac{\gamma}{\|\tilde D\|_2+\varepsilon_\kappa}\,\tilde D . \label{eq:D}
\end{align}
Then define
\begin{align}
&W=\begin{bmatrix}P&PA\\ A^*P&P\end{bmatrix},\qquad
Z=\begin{bmatrix}\gamma I_{n_d}&D^*\\ D&\gamma I_{n_z}\end{bmatrix},\\
&\mathcal M=
\begin{bmatrix}
\mathds{1}_{n_h\times n_d} & 0_{n_h\times n_z} \\
0_{n_h\times n_d} & \mathds{1}_{n_h\times n_z}
\end{bmatrix},\\
&\tilde Y=\mathcal M\odot\bar Y,\qquad
\eta=1+\left\|W^{-1/2}\tilde Y Z^{-1/2}\right\|_2,\\
&Y=\eta^{-1}\tilde Y=\begin{bmatrix}Y_{21}&0\\0&Y_{22}\end{bmatrix}.
\end{align}
Finally, set
\begin{align}
    B=P^{-1}Y_{21},\qquad C=Y_{22}^* .
\end{align}
\end{mybox}

\begin{theorem}\label{thm:param_suff}
The map $\kappa_\gamma$ defined in Block~\ref{box3} is a free parametrization of $\mc G_\gamma$.
\end{theorem}

\begin{proof}
The proof is based on the Schur complement. We construct $A,B,C,D$ so that the bounded-real LMI
\begin{equation}
\label{eq:real_bounded_lemma_discrete}
\Gamma:=
\begin{bmatrix}
P & PA & PB & 0 \\
* & P & 0 & C^* \\
* & * & \gamma I_{n_d} & D^* \\
* & * & * & \gamma I_{n_z}
\end{bmatrix}
\succ0
\end{equation}
holds. Satisfaction of~\eqref{eq:real_bounded_lemma_discrete} implies that the corresponding DT LTI system has $\mc L_2$-gain not larger than $\gamma$.

Partition
\[
\Gamma=\begin{bmatrix}W&Y\\Y^*&Z\end{bmatrix},
\]
with $W,Y,Z$ as in Block~\ref{box3}. By construction,
$|\lambda_j|=\exp(-\exp(\mu_j))<1$, so $A$ is Schur stable. Since $P=A^*A+\varepsilon_\kappa I_{n_h}\succ0$ and $A$ commutes with $P$,
\[
P-A^*PA
=\operatorname{diag}\!\left((|\lambda_j|^2+\varepsilon_\kappa)(1-|\lambda_j|^2)\right)\succ0.
\]
By the Schur complement, $W\succ0$. Moreover,
\[
\|D\|_2=\frac{\gamma\|\tilde D\|_2}{\|\tilde D\|_2+\varepsilon_\kappa}<\gamma,
\]
and therefore $Z\succ0$.
Finally, the scaling by $\eta$ gives
\[
\left\|W^{-1/2}YZ^{-1/2}\right\|_2<1,
\]
which implies $Y^*W^{-1}Y\prec Z$. Applying the Schur complement to $\Gamma$ yields $\Gamma\succ0$. Since the construction is defined for all free parameters, except for the null sets already allowed in Definition~\ref{def:param}, $\kappa_\gamma$ is a free parametrization of $\mc G_\gamma$.
\end{proof}

The comments on spectral-norm computation and on making $\gamma$ trainable are the same as in Remarks~\ref{rem:complexity} and~\ref{rem:bound}. 

\subsubsection*{Initialization and computational efficiency}
One advantageous aspect of the parametrization $\kappa_{\gamma}$ is the decoupling between the parametrization of the matrix $A$ and that of the remaining matrices. As shown in~\eqref{eq:A1} and~\eqref{eq:A2}, we adopt the complex-diagonal parametrization introduced in~\cite{orvieto_resurrecting_2023}, which directly represents the eigenvalues of $A$ in the complex plane through their modulus and phase. This structure greatly simplifies the long-memory initialization described in the previous section, as the modulus of each eigenvalue can be set arbitrarily close to the unit circle. Moreover, the phase of each eigenvalue can be tuned independently, an adjustment that, as discussed in~\cite{orvieto_resurrecting_2023}, may improve training stability when small phase values are enforced. Finally, the complex-diagonal parametrization of the matrix $A$ makes time-domain simulation highly efficient, as algorithms such as parallel scan can exploit such a structure to parallelize computations, substantially reducing the time required to generate trajectories compared to standard recursive simulation \cite{blelloch_prefix_2004}.

\subsection{L2RU scaffolding: full prescribed-bound parametrization}

We now combine the LTI cores with the nonlinear and linear components of the SSM, to form the L2RU architecture. Let $\chi_\gamma$ denote either the projected square parametrization $\pi\circ\psi_\gamma$ or the structured parametrization $\kappa_\gamma$. Let $\mu_{\nu,\zeta}$ be a freely parametrized family of pointwise nonlinearities with $\mu_{\nu,\zeta}(0)=0$ and Lipschitz bound $\zeta$.

\begin{mybox}{L2RU parametrization}{box2}
Given $\hat\gamma>0$ and free parameters
\begin{align*}
   \left\{\{\omega_i,\tilde\gamma_i,\nu_i,\tilde\zeta_i\}_{1\le i\le r},\tilde E,\tilde H\right\},
\end{align*}
define $\rho_{\hat\gamma}$ as follows. For $1\le i\le r$, set
\begin{align}
&\gamma_i=|\tilde\gamma_i|,\qquad \zeta_i=|\tilde\zeta_i|,\\
&(A_i,B_i,C_i,D_i)=\chi_{\gamma_i}(\omega_i), \label{eq:psys}\\
&\mu_i=\mu_{\nu_i,\zeta_i}.
\end{align}
For the encoder and decoder, set
\begin{align}
&E=\tilde E, \label{eq:pE}\\
&H=\frac{\tilde H\hat\gamma}{\|\tilde H\|_2\|\tilde E\|_2}
\prod_{i=1}^{r}\left(\gamma_i\zeta_i+1\right)^{-1}. \label{eq:pF}
\end{align}
The map is defined almost everywhere, excluding the null set where $\tilde E=0$ or $\tilde H=0$.
\end{mybox}

\begin{theorem}\label{thm:param2}
The map $\rho_{\hat\gamma}$ defined in Block~\ref{box2} is a free parametrization of $\Theta_{\hat\gamma}$.
\end{theorem}

\smallskip
Theorem \ref{thm:param2}, proved in \cite{massai_free_2025},  provides a straightforward method to freely parameterize all the individual components of the model such that $f_{\rho_{\hat{\gamma}}}$ maintains a fixed $\mc L_2$-bound $\hat{\gamma}$. It is worth noting that the $\mc L_2$-bounds and Lipschitz constants of the linear systems and nonlinearities in each layer are treated as free parameters in the definition of $\rho_{\gamma}$ \footnote{Technically $\gamma, \zeta$ are not free parameters as they must be positive, however, we can simply set $\gamma=|\tilde{\gamma}|, \zeta=|\tilde{\zeta}|$ where $\tilde{\gamma}, \tilde{\zeta}$ are free parameters (or, equivalently, use any smooth positive reparametrization).}. Their values need not be fixed a priori; instead, they are used to properly normalize the $\mc L_2$-bound of the decoder $H$, to guarantee that the overall bound of the L2RU architecture equals the prescribed $\gamma$. The key point is that this requires explicitly the $\mc L_2$-bounds and Lipschitz constants, which is made possible by the parametrization $\kappa_{\gamma}$ for the linear systems and by the definition of $\mu_{\xi}$ for the nonlinearities.

\section{Comparison of the proposed L2RU parametrizations}
\label{sec:comparison}

We now compare the two L2RU variants induced by the parametrizations $\psi_{\gamma}$ and $\kappa_{\gamma}$. Both constructions provide free parametrizations of LTI systems with prescribed $\mc L_2$-bound and, therefore, they can be used to induce the full L2RU parametrization via Theorem~\ref{thm:param2}. However, they make fundamentally different compromises. The parametrization $\psi_{\gamma}$ is designed to maximize expressiveness and certificate tightness, whereas $\kappa_{\gamma}$ is designed to preserve the diagonal structure that makes modern SSMs computationally efficient.

The comparison is first and foremost a comparison between the LTI cores generated by $\psi_{\gamma}$ and $\kappa_{\gamma}$. This distinction is important because the full L2RU model contains additional components, namely encoder and decoder maps, Lipschitz nonlinearities, and skip connections. These components are essential for building expressive SSMs, but they can also partially obscure the effect of the underlying recurrent parametrization. For instance, in a full nonlinear SSM, the nonlinearities or the encoder-decoder maps may compensate for limitations of the LTI core, making it difficult to attribute performance differences solely to $\psi_{\gamma}$ or $\kappa_{\gamma}$.

For this reason, the theoretical comparison focuses on the LTI cores, where precise statements can be made about completeness, conservatism, gain tightness, initialization, and scan compatibility. The empirical comparison mirrors this logic. We first compare the bare LTI cores in isolation, so that the differences between $\psi_{\gamma}$ and $\kappa_{\gamma}$ are not hidden by the surrounding SSM scaffolding. We then embed both cores into the same full L2RU architecture, with identical encoder, decoder, nonlinearities, skip connections, and training protocol, to assess whether the core-level differences identified theoretically remain visible in the complete model.

In the following, we refer to an L2RU whose LTI blocks are parametrized by $\psi_{\gamma}$ as a $\psi_{\gamma}$-L2RU, and to one whose LTI blocks are parametrized by $\kappa_{\gamma}$ as a $\kappa_{\gamma}$-L2RU. When discussing completeness, conservatism, gain tightness, initialization, and scan compatibility, the comparison should be understood at the level of the LTI core unless explicitly stated otherwise. For LTI systems, the induced $\mc L_2$-gain coincides with the discrete-time $H_\infty$ norm; in the experiments below, we therefore use the $H_\infty$ norm only to evaluate the actual gain of target and learned LTI systems numerically.

\renewcommand{\arraystretch}{1.25}
\begin{table*}[t]
\centering
\caption{Comparison between the LTI cores generated by $\psi_{\gamma}$ and $\kappa_{\gamma}$, and their implications for the corresponding L2RU architectures. Parameter counts are reported for square cores of dimension $n$.}
\rowcolors{2}{gray!10}{white}
\begin{tabular}{@{}p{4.2cm}p{5.1cm}p{5.1cm}@{}}
\toprule
\textbf{Property} & \textbf{$\psi_{\gamma}$-L2RU} & \textbf{$\kappa_{\gamma}$-L2RU} \\
\midrule
Guaranteed prescribed $\mc L_2$-bound
& \cmark
& \cmark \\

LTI-core completeness
& Complete for square systems, up to null sets
& Not complete; based on sufficient conditions \\

Systems handled directly
& Square LTI cores; arbitrary external dimensions through encoder/decoder
& General, possibly non-square LTI cores \\

Tightness of prescribed gain
& Can represent systems with gain arbitrarily close to $\gamma$
& Certified upper bound, but tightness is not guaranteed in general \\

State matrix structure
& Dense real matrix
& Complex diagonal matrix \\

Parallel scan compatibility
& \xmark
& \cmark \\

Eigenvalue initialization
& Isotropic pole modulus controlled by Proposition~\ref{prop:init}; phase through $Q$
& Independent control of modulus and phase through~\eqref{eq:A1}--\eqref{eq:A2} \\

Number of effective real parameters
& $\tfrac{1}{2}(9n^2+n)+2$
& $6n^2+2n$ \\

Main advantage
& Expressiveness and tight certificates
& Scan efficiency and simple spectral initialization \\

Main limitation
& Sequential simulation; square LTI cores
& Conservatism and possible loss of expressiveness \\
\bottomrule
\end{tabular}
\rowcolors{2}{white}{white}
\label{tab:comparison}
\end{table*}

\subsection{Expressiveness and certificate tightness}

The main advantage of the $\psi_{\gamma}$ core is completeness. By Theorem~\ref{thm:param}, $\psi_{\gamma}$ provides a free and complete parametrization of $\mc R_{\gamma}^n$, and therefore of $\mc G_{\gamma}^n$ through the projection map~\eqref{eq:proj}. Hence, up to the null sets allowed in Definition~\ref{def:param}, every square LTI system satisfying the prescribed $\mc L_2$-bound is represented. In particular, $\psi_{\gamma}$ does not lose systems whose true induced $\mc L_2$-gain approaches the prescribed value $\gamma$.

As anticipated in the previous section, the requirement that the LTI core be square should be understood only as an internal structural constraint of the parametrization. In the full L2RU architecture, the external input and output dimensions $n_u$ and $n_y$ are connected to the state-space block through the encoder and decoder in~\eqref{eq:l2ruseries}. The square size $n$, corresponding to $n_h=n_d=n_z=n$, is therefore an internal width parameter, chosen independently of the data dimensions. Thus, the square-core assumption is not a restriction on the input-output dimensions of the learned operator, but rather the condition under which the complete parametrization $\psi_{\gamma}$ applies.

This distinction is relevant when $\gamma$ is used as an actual design parameter. In system identification or robust control, for example, one may want to impose a gain budget that is tight enough to be useful while still guaranteed throughout training. Completeness ensures that, within the square LTI setting, the parametrization does not itself exclude admissible systems near the boundary of the prescribed certificate. Consequently, if a target system with gain close to $\gamma$ is hard to learn with $\psi_{\gamma}$, the limitation should be sought in optimization, initialization, or data availability, rather than in a loss of coverage of the bounded-real set.

The parametrization $\kappa_{\gamma}$ has a different nature. Theorem~\ref{thm:param_suff} guarantees that every LTI core generated by $\kappa_{\gamma}$ has $\mc L_2$-gain at most $\gamma$, but the construction is based on sufficient conditions for the bounded-real inequality. Hence, $\kappa_{\gamma}$ is free but not complete. It may generate systems with gain close to $\gamma$, but this is not guaranteed for all bounded systems. Consequently, poor fit of the bare $\kappa_{\gamma}$ core may reflect either optimization difficulty or intrinsic conservatism of the parametrization.

This distinction is central for interpreting numerical results. A larger prescribed value of $\gamma$ enlarges the set of systems satisfying the desired $\mc L_2$-bound, but it does not necessarily imply that a conservative parametrization explores that set effectively. Thus, the relevant empirical question is not only whether the learned system satisfies the bound, but also whether the parametrization can use the available gain budget when the task requires it.

\paragraph{Experiment 1: Bare-core LTI identification and gain-budget usage.}
To isolate the expressiveness of the two recurrent parametrizations, we first consider a purely linear system-identification task. In this experiment there are no encoder or decoder maps, no nonlinearities, and no skip connections. The learned model is only the LTI core generated either by $\psi_{\gamma}$ or by $\kappa_{\gamma}$. This setting is therefore the empirical counterpart of the core-level comparison in Table~\ref{tab:comparison}.

The target system is a dense stable DT LTI system
\begin{align}
G_\star :=
\left(
A_\star,B_\star,C_\star,D_\star
\right),
\end{align}
with input and output dimensions equal to the core dimension $n$. Unless otherwise stated, we use $n=8$. The state matrix is generated from real rotation blocks and then densified through an orthogonal similarity transformation. More precisely, we first construct a block-diagonal matrix $\bar A_\star$ whose $2\times2$ blocks are of the form
\begin{align}
    r_j
    \begin{bmatrix}
        \cos(\omega_j) & -\sin(\omega_j)\\
        \sin(\omega_j) & \cos(\omega_j)
    \end{bmatrix},
\end{align}
with radii $r_j\in[0.72\rho_\star,\rho_\star]$ and frequencies $\omega_j\in[0.15\pi,0.85\pi]$. We then set
\begin{align}
    A_\star = Q_\star \bar A_\star Q_\star^\top,
\end{align}
where $Q_\star\in\mc S\mc O(n)$ is sampled randomly. This gives a dense stable target with spectral radius at most $\rho_\star$. The matrices $B_\star,C_\star,D_\star$ are sampled from Gaussian distributions and then jointly rescaled so that
\begin{align}
    \|G_\star\|_\infty=\gamma_\star .
\end{align}
In the reported experiment we use $\rho_\star=0.9$ and $\gamma_\star=3$. The value $\gamma_\star$ is marked in the plots as the target gain.

Training and validation data are generated by exciting $G_\star$ with i.i.d. Gaussian inputs,
\begin{align}
    u_t \sim \mc N(0,I_n),
\end{align}
and simulating the target from zero initial state. We use $N_{\rm tr}=64$ training trajectories and $N_{\rm val}=64$ validation trajectories, each of length $T=200$. The same target system and datasets are used for all values of $\gamma$ and for both parametrizations.

For each prescribed bound
\begin{align}
    \gamma \in \{1,2,2.5,3,4,5,6,8,12\},
\end{align}
we train a $\psi_{\gamma}$ core and a $\kappa_{\gamma}$ core by minimizing the trajectory MSE. 
After training, we extract the learned state-space matrices and compute the learned $H_\infty$ norm on a uniform frequency grid over $[0,\pi]$. Since the systems are LTI, this coincides with the induced $\mc L_2$-gain. All curves report mean and standard deviation over five random initializations.

\begin{figure*}[t]
\centering
\includegraphics[scale=.4]{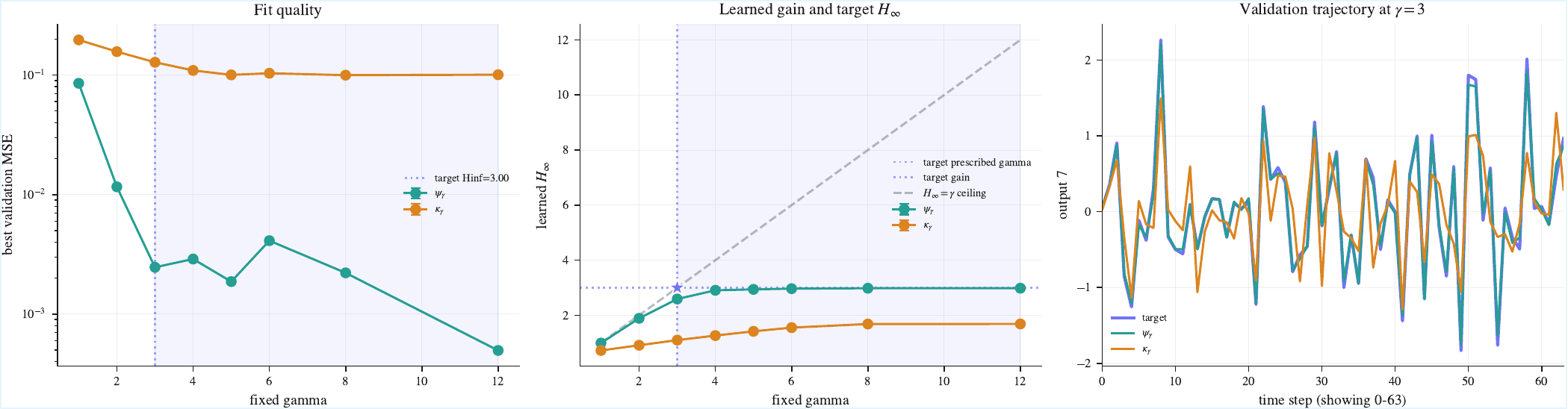}
\caption{Bare-core LTI identification and gain-budget usage. The experiment isolates the LTI cores generated by $\psi_{\gamma}$ and $\kappa_{\gamma}$, without nonlinearities, encoder-decoder maps, or skip connections.}
\label{fig:bare_core_fit}
\end{figure*}

The two cores are initialized according to their natural parametrizations. For $\psi_{\gamma}$, we use the long-memory initialization of Proposition~\ref{prop:init}, with the common initial pole radius set close to the unit circle. For $\kappa_{\gamma}$, the diagonal poles are initialized directly in the complex plane, with radii sampled over a broad stable range and phases sampled over $[0,\pi]$. In both cases, the prescribed bound $\gamma$ is kept fixed during training. The implementation details are summarized in Table~\ref{tab:exp1_setup}, and the results are reported in Fig.~\ref{fig:bare_core_fit}. The left panel shows the best validation MSE obtained for each prescribed value of $\gamma$. Across the entire range of prescribed bounds, the $\psi_{\gamma}$ core achieves substantially lower MSE than the $\kappa_{\gamma}$ core. This difference is also visible in the right panel, which shows a representative validation trajectory: the trajectory generated by $\psi_{\gamma}$ tracks the target system more accurately than the one generated by $\kappa_{\gamma}$.

The central panel reports the learned $\mc L_2$-gain, computed as the discrete-time $H_\infty$ norm of the learned LTI system, as a function of the prescribed bound $\gamma$. As expected, both parametrizations always produce systems whose learned gain remains below the prescribed value, confirming the validity of the certificate. The behavior of the two parametrizations, however, is markedly different. The $\psi_{\gamma}$ core is able to recover the true gain $\gamma_\star=3$ of the target system when the prescribed budget is sufficiently loose. This is consistent with the completeness of $\psi_{\gamma}$, which allows systems with gain arbitrarily close to the prescribed bound to be represented.

At the same time, the experiment also illustrates a numerical subtlety of tight certificates. Although $\psi_{\gamma}$ can theoretically represent systems with gain arbitrarily close to $\gamma$, the parametrization becomes increasingly ill-conditioned near the boundary of the certified set. This explains why, when the prescribed value is exactly $\gamma=\gamma_\star=3$, the learned gain remains slightly below the target gain, whereas prescribing a moderately looser bound allows the model to recover $\gamma_\star$ accurately. Moreover, the first panel shows that both parametrizations tend to achieve lower validation error as the prescribed budget $\gamma$ is relaxed, with the effect being particularly pronounced for $\psi_{\gamma}$. This is consistent with the fact that a looser gain constraint enlarges the feasible set and makes the optimization landscape less restrictive.
 By contrast, the $\kappa_{\gamma}$ core remains noticeably more conservative: even when larger values of $\gamma$ are prescribed, the learned gain stays well below the target gain. This behavior is consistent with the fact that $\kappa_{\gamma}$ is based on sufficient, rather than complete, bounded-real conditions.

\begin{table}[t]
\centering
\caption{Setup for Experiment 1.}
\label{tab:exp1_setup}
\footnotesize
\begin{tabular}{@{}p{0.43\columnwidth}p{0.49\columnwidth}@{}}
\toprule
Quantity & Value \\
\midrule
Task & Bare-core LTI system identification \\
Target dimension & $n_\star=8$ \\
Model input/output dimension & $n_d=n_z=8$ \\
Target spectral radius & $\rho_\star=0.9$ \\
Target gain & $\|G_\star\|_\infty=\gamma_\star=3$ \\
Training trajectories & $N_{\rm tr}=64$ \\
Validation trajectories & $N_{\rm val}=64$ \\
Trajectory length & $T=200$ \\
Input distribution & $u_t\sim\mc N(0,I_8)$ \\
Prescribed gains & $\{1,2,2.5,3,4,5,6,8,12\}$ \\
Optimizer & Adam \\
Learning rate & $10^{-3}$ \\
Max epochs & $2000$ \\
Early-stopping patience & $300$ epochs \\
Seeds & $5$ \\
$H_\infty$ grid & $4096$ frequencies in $[0,\pi]$ \\
$\psi_{\gamma}$ rollout & Dense recurrent loop \\
$\kappa_{\gamma}$ rollout & Parallel scan \\
\bottomrule
\end{tabular}
\end{table}

\subsection{State-space structure and computational efficiency}

The price paid by $\psi_{\gamma}$ for completeness is structural. The state matrix $A$ generated by~\eqref{eq:pari} is a dense real matrix. As a result, the associated recurrence must be simulated sequentially through the standard state update. This is appropriate for moderate sequence lengths and for settings where certificate tightness is the main objective, but it does not exploit fast parallel scan algorithms that take advantage of diagonal or block-diagonal structures.

By contrast, $\kappa_{\gamma}$ preserves the complex-diagonal representation
\begin{align}
    A=\mathrm{diag}(\lambda_1,\ldots,\lambda_{n_h}),
\end{align}
with the eigenvalues directly parametrized by~\eqref{eq:A1}. This is the same structural principle used in LRU-type SSMs and makes the recurrence compatible with parallel scan implementations. Therefore, $\kappa_{\gamma}$ is expected to be significantly faster on long sequences, especially when the sequence length is the computational bottleneck.

Thus, the comparison is not simply between a ``better'' and a ``worse'' parametrization. Instead, $\psi_{\gamma}$ and $\kappa_{\gamma}$ occupy different points on the expressiveness-efficiency tradeoff. The parametrization $\psi_{\gamma}$ provides a complete square parametrization and can represent systems arbitrarily close to the prescribed-bound boundary, but it produces dense recurrences. The parametrization $\kappa_{\gamma}$ sacrifices completeness to retain the scan-compatible diagonal structure that is essential in many long-sequence SSM applications.

\paragraph{Experiment 2: Runtime scaling in sequence length and state dimension.}

We compare the simulation cost of the dense $\psi_{\gamma}$ core with the cost of the $\kappa_{\gamma}$ core simulated either recursively or through parallel scan. This experiment is not intended to evaluate prediction accuracy. Its purpose is to isolate the computational consequence of the state-space structure discussed above: $\psi_{\gamma}$ produces a dense real state matrix and is therefore evaluated through a standard recurrent loop, whereas $\kappa_{\gamma}$ preserves a complex-diagonal state matrix and can exploit scan-based simulation.

The compared implementations are: (i) the $\psi_{\gamma}$ core simulated through the dense recurrent loop; (ii) the $\kappa_{\gamma}$ core simulated through the same sequential recurrent loop; and (iii) the $\kappa_{\gamma}$ core simulated using parallel scan. Thus, comparing (i) and (ii) isolates the cost of the two parametrizations when both are evaluated sequentially, whereas comparing (ii) and (iii) isolates the computational advantage enabled by the scan-compatible structure of $\kappa_{\gamma}$.

We report forward-plus-backward wall-clock time, since this is the relevant cost during training. All timings use random input batches
$    u \in \mathbb{R}^{N_b \times T \times n},$
with batch size $N_b=4$, prescribed gain $\gamma=3$, and are averaged over 10 repetitions after 3 warm-up runs. The experiment contains two sweeps. First, we fix the square core dimension to $n=8$ and vary the sequence length as
\[
    T \in \{32,64,128,256,512,1024,2048\}.
\]
Second, we fix the sequence length to $T=512$ and vary the square core dimension as
\[
    n \in \{4,8,16,32,64,128,256\}.
\]
The second sweep is included to separate temporal scaling from state-dimension scaling. The number of free parameters grows quadratically with $n$ for both parametrizations, namely $\frac{1}{2}(9n^2+n)+2$ for $\psi_{\gamma}$ and $6n^2+2n$ for $\kappa_{\gamma}$.

\begin{figure}[t]
\centering
\includegraphics[scale=.288]{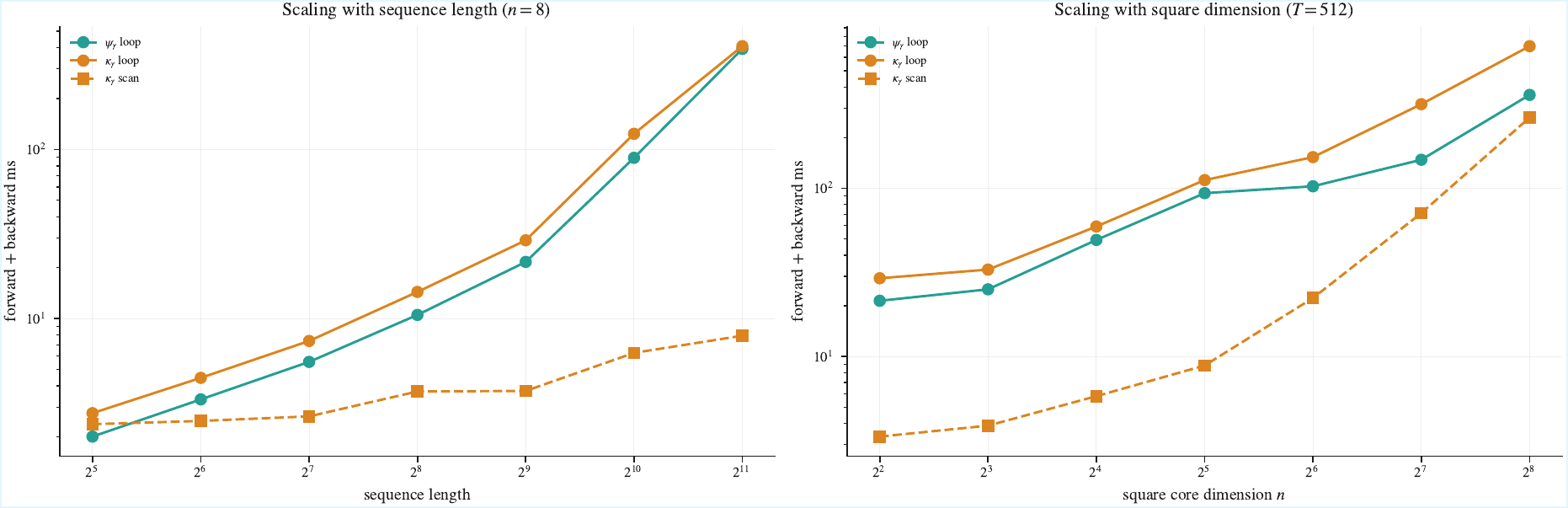}
\caption{Runtime comparison. The scan-compatible structure of $\kappa_{\gamma}$ is expected to provide an advantage for long sequences, while $\psi_{\gamma}$ requires dense recursive simulation.}
\label{fig:runtime_scan}
\end{figure}

The results are shown in Fig.~\ref{fig:runtime_scan}. The left panel confirms the expected advantage of the scan-compatible implementation along the sequence-length direction. The two loop-based implementations have comparable scaling in $T$, while the scan implementation of $\kappa_{\gamma}$ becomes increasingly advantageous as the sequence length grows. This is consistent with the role of associative scan algorithms: a recurrent loop has sequential depth $O(T)$, whereas a parallel scan reduces the sequential depth to $O(\log T)$ while retaining linear total work in $T$. For short sequences, the scan overhead is not fully amortized; for long sequences, the reduced sequential depth becomes the dominant effect.

The right panel gives a more nuanced picture. When $T$ is fixed and the state dimension $n$ increases, the scan advantage is less pronounced. This is expected: scan parallelizes the temporal recurrence, but it does not remove the cost associated with the state dimension, the parameterization maps, or the memory traffic of the backward pass. Consequently, $\kappa_{\gamma}$ should not be interpreted as uniformly faster in every computational regime. Its advantage is most clearly a long-sequence advantage. In regimes where the state dimension is large and the horizon is moderate, the dense $\psi_{\gamma}$ loop can remain competitive, despite lacking scan compatibility.

Overall, the experiment supports the theoretical tradeoff in Table~\ref{tab:comparison}. The $\kappa_{\gamma}$ parametrization is preferable when long-sequence throughput is the main bottleneck, because its diagonal structure enables parallel scan. The $\psi_{\gamma}$ parametrization sacrifices this algorithmic advantage in exchange for completeness and tighter use of the prescribed gain budget.

\subsection{Initialization and long-memory behavior}

Both parametrizations support long-memory initialization, but through different mechanisms. For $\psi_{\gamma}$, Proposition~\ref{prop:init} gives the eigenvalue modulus at initialization in closed form:
\begin{align}
    r_0:=|\lambda|=\sqrt{\frac{2\sigma(\alpha)}{3-\sigma(\alpha)}}.
\end{align}
Choosing $\alpha$ so that $\sigma(\alpha)\to 1$ places the initial poles arbitrarily close to the unit circle while preserving the $\mc L_2$ certificate. The limiting case yields
\begin{align}
    A_0=r_0Q,
\end{align}
with $Q$ orthogonal: every eigenvalue shares the same modulus $r_0$, and the phases are determined by $Q$ via~\eqref{eq:orth}. Taking $Q$ as a block-diagonal rotation prescribes those phases explicitly, but $\alpha$ controls only a single, common memory timescale, not the modulus of each individual pole.

By contrast, the diagonal parametrization~\eqref{eq:A1} of $\kappa_{\gamma}$ exposes both modulus and phase of each mode independently. Poles can therefore be initialized close to the unit circle, distributed over a chosen radius range, and given small phases as in LRU-type architectures \cite{orvieto_resurrecting_2023}.

\begin{figure*}[t]
\centering
\includegraphics[scale=.4]{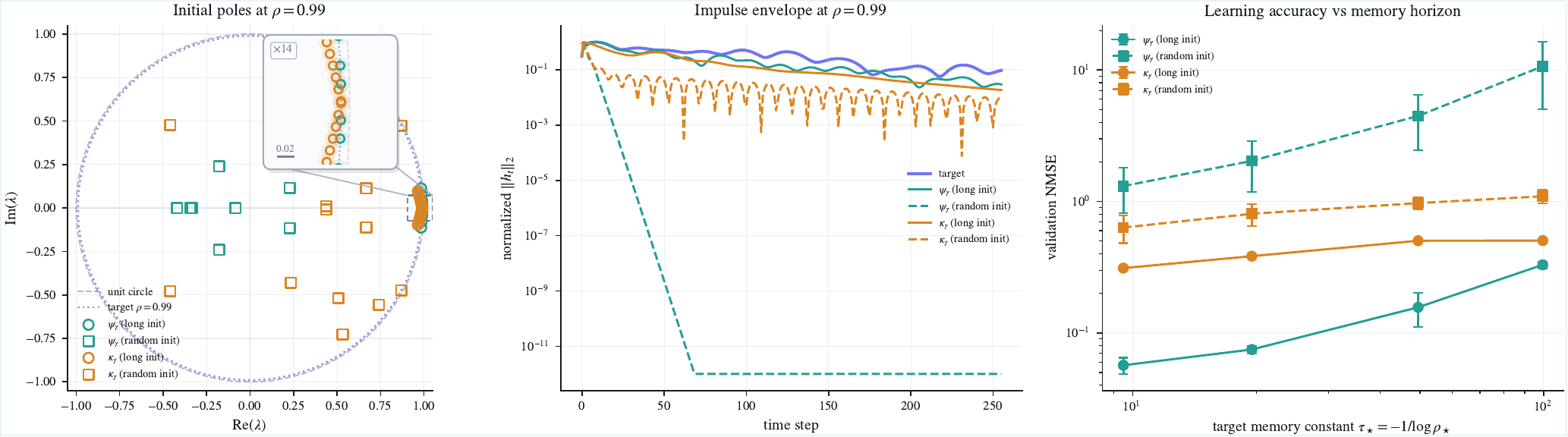}
\caption{\textbf{Long-memory identification and initialization sensitivity.}
Bare $\psi_\gamma$- and $\kappa_\gamma$-cores with prescribed bound $\gamma=3$ are trained to identify dense target LTI systems with fixed induced gain $\|G_\star\|_\infty=3$ and increasing memory horizon. \emph{Left:} initialized poles for a representative hard target with $\rho_\star=0.995$, shown together with the unit circle and the circle of radius $\rho_\star$. \emph{Center:} normalized impulse envelope $e_t=\|h_t\|_2/\max_s\|h_s\|_2$ for the same target. \emph{Right:} best validation NMSE versus the target memory constant $\tau_\star=-1/\log\rho_\star$, computed on full validation trajectories and normalized by the corresponding output power. Structured long-memory initialization improves both parametrizations, especially $\psi_\gamma$, while $\kappa_\gamma$ is more robust in the most persistent regime.}
\label{fig:full_ssm_comparison}
\end{figure*}

\paragraph{Experiment 3: Initialization geometry and long-memory identification.}
We probe this distinction in two stages. The first visualizes the spectral geometry: $\psi_{\gamma}$ initializations $A_0=r_0Q$ place all poles on a common circle of radius $r_0$ with phases set by $Q$, whereas $\kappa_{\gamma}$ assigns each mode its own radius and phase. The pole panel of Fig.~\ref{fig:full_ssm_comparison} contrasts the two patterns directly.

The second stage tests whether this geometric flexibility translates into a learning advantage. We consider dense target LTI systems with fixed gain
\begin{align}
    \gamma_\star=3
\end{align}
and increasing memory horizon. Learned models are bare LTI cores of state dimension $n=8$; targets have state dimension $16$. For each
\begin{align}
    \rho_\star \in \{0.9,\,0.95,\,0.98,\,0.99,\,0.995\},
\end{align}
we draw $4$ blueprints by forming a real block-diagonal matrix with eigenvalues on the unit circle and small phases $\phi_j \in [-0.05\pi,\;0.05\pi]$, then applying an orthogonal similarity transform to obtain a dense $A_{\mathrm{shape}}$. Scaling
\begin{align}
    A_\star=\rho_\star A_{\mathrm{shape}}
\end{align}
sets the memory horizon, and the input, output, and feedthrough matrices are rescaled per blueprint so that $\gamma_\star=3$. Varying $\rho_\star$ thus moves the memory timescale at fixed gain.
We index results by the equivalent memory constant
$    \tau_\star:=-\frac{1}{\log \rho_\star},$
which directly measures the dominant decay timescale, a larger $\tau_\star$ means longer memory.

For each blueprint we generate $8$ training and $8$ validation trajectories of length $64$ from i.i.d.\ Gaussian inputs. Both parametrizations train at $\gamma=3$ for up to $500$ epochs with early-stopping patience $80$. Combined with $3$ seeds per configuration, each aggregate point averages $12$ independent runs.

We compare a structured long-memory initialization against a random one for each model. The structured $\psi_{\gamma}$ uses the isotropic eye initialization of Proposition~\ref{prop:init}; its random variant initialize each free parameter randomly. The structured $\kappa_{\gamma}$ places diagonal poles near the target memory scale; its random variant draws radii from the open unit disk and phases from the upper half-circle. It is worth noting that the two random schemes are native to their respective parametrizations and thus do not yield, in general, the same eigenvalue distribution over the unit circle.

Figure~\ref{fig:full_ssm_comparison} reports the results. The \emph{left panel} compares the pole locations induced by the different initialization schemes. The unit circle marks the stability boundary, while the circle of radius $\rho_\star=0.995$ indicates the memory scale of the target dynamics. Since poles close to this reference circle produce slowly decaying responses, an initialization concentrated near it provides a suitable long-memory prior.

The structured $\kappa_{\gamma}$ initialization places the diagonal modes directly in this region, with radii and phases chosen mode by mode. The structured $\psi_{\gamma}$ initialization instead produces a ring of poles with common modulus $\rho_\star$, reflecting the isotropic-radius geometry of the construction. In contrast, the random initializations are not adapted to the target memory scale: their poles are spread without regard to the desired radius, making the initial dynamics less persistent and therefore less aligned with the system to be identified.

The \emph{central panel} shows the normalized impulse envelope for the same target. We excite the first input channel with a unit impulse, denote the resulting output vector at time $t$ by $h_t$, and plot $e_t := \|h_t\|_2 / \max_s \|h_s\|_2$. This normalization removes the overall response scale and isolates the temporal decay profile: slowly decaying envelopes correspond to long-memory dynamics, while rapidly decaying envelopes correspond to short-memory dynamics. For each model class, we display the run with the lowest validation NMSE on a representative target realization with $\rho_\star=0.995$. Both structured initializations preserve the slow decay much more faithfully than their random counterparts. The effect is most pronounced for random $\psi_{\gamma}$, whose response quickly collapses to short-memory behavior, whereas structured $\kappa_{\gamma}$ gives the closest match to the target envelope.

The \emph{right panel} reports the best validation NMSE as a function of the target memory constant $\tau_\star$. Let $\{(u^{(i)}_{0:T-1},y^{(i)}_{0:T-1})\}_{i=1}^{N_{\mathrm{val}}}$ be the validation set, with $N_{\mathrm{val}}=8$ and $T=64$. At each epoch, we compute
\begin{align}
    \mathrm{MSE}_{\mathrm{val}}
    =
    \frac{1}{N_{\mathrm{val}}Tn_y}
    \sum_{i=1}^{N_{\mathrm{val}}}
    \sum_{t=0}^{T-1}
    \|\hat y_t^{(i)}-y_t^{(i)}\|_2^2 .
\end{align}
The plotted quantity is the best value reached during training, normalized by the validation output power:
\begin{align}
    \mathrm{NMSE}_{\mathrm{val}}
    =
    \frac{
        \min_{\mathrm{epoch}} \mathrm{MSE}_{\mathrm{val}}
    }{
        \frac{1}{N_{\mathrm{val}}Tn_y}
        \sum_{i=1}^{N_{\mathrm{val}}}
        \sum_{t=0}^{T-1}
        \|y_t^{(i)}\|_2^2
    } .
\end{align}
We use NMSE because all target systems are rescaled to the same induced $\mc L_2$ gain, but this does not equalize their average finite-horizon output power under Gaussian inputs. Changing the pole radius changes how the gain is distributed across frequencies, and therefore changes the typical energy of the validation trajectories. The normalization makes the comparison across different $\tau_\star$ values reflect relative prediction accuracy rather than output scale.

As $\tau_\star$ increases, the target dynamics become more persistent and successful identification requires useful poles near the unit circle. The results show that random initialization is particularly harmful for $\psi_{\gamma}$ in this regime, whereas its structured initialization remains much more effective. For $\kappa_{\gamma}$, the gap between random and structured initialization is milder, consistent with the direct spectral access provided by the diagonal parametrization. Overall, and coherently with the previous findings,  $\psi_{\gamma}$ still performs better in terms of learning accuracy, provided that a good spectral initialization is enforced. In the end, the experiment shows that long-memory performance depends not only on the expressiveness of the parametrization, but also on whether its initialization geometry places the model in a suitable dynamical regime.

\subsection{Implications for learning}

The theoretical comparison above concerns the LTI cores in isolation, whereas in the full L2RU architecture the recurrent block is embedded into a larger nonlinear scaffold. Encoder and decoder maps, static nonlinearities, and residual connections can partially compensate for limitations of the core and may therefore blur the direct effect of completeness or conservatism on the final prediction error. At the same time, the LTI core remains the component that provides the recurrent dynamics and the associated $\mathcal L_2$ certificate. Its parametrization is therefore still expected to matter in regimes where gain usage, memory, and recurrent expressiveness are important.

This motivates a two-level empirical comparison. The bare-core experiments isolate the recurrent parametrizations themselves. The full-SSM experiments then ask whether the same differences remain visible once both cores are placed inside the same nonlinear architecture and, more broadly, whether the resulting L2RU models are competitive trainable sequence models on nonlinear learning tasks. Since the purpose of this benchmark is predictive performance rather than robustness under a prescribed gain budget, we keep the bound $\gamma$ trainable. This prevents the comparison from being dominated by a manually chosen value of $\gamma$ and lets both parametrizations use the gain level favored by the data and optimization procedure.

\paragraph{Experiment 4: Full-SSM same-scaffold comparison.}
We compare $\psi_{\gamma}$ and $\kappa_{\gamma}$ inside the same L2RU architecture on the \emph{Cascaded Tanks} benchmark from the \emph{nonlinearbenchmark.org} collection,\footnote{\url{https://www.nonlinearbenchmark.org/benchmarks/cascaded-tanks}. The corresponding Python dataloader is available in the \texttt{nonlinear\_benchmarks} repository: \url{https://github.com/GerbenBeintema/nonlinear_benchmarks}.} a standard nonlinear system-identification problem based on a two-tank fluid-level system with overflow. The official dataset provides one estimation trajectory and one test trajectory
with a benchmark-specified warm-up window of $50$ samples. All reported RMSE values are computed after discarding this initial segment.

Both models use the same three-layer L2RU scaffold with internal width $d_{\mathrm{model}}=8$, learned initial state, and trainable global gain initialized at $\gamma=1$. The nonlinear blocks are GLU feedforward modules with hidden width $d_{\mathrm{hidden}}=12$ and three nonlinear layers. Training is performed for $1000$ epochs with Adam and learning rate $2\cdot 10^{-2}$. To keep the comparison fair, we match the parameter budgets as closely as possible by using state dimension $8$ for $\psi_{\gamma}$ and $7$ for $\kappa_{\gamma}$, yielding $1358$ and $1385$ trainable parameters, respectively.

Results are reported in Table~\ref{tab:full_ssm_tanks_best}. The $\psi_{\gamma}$-based model achieves lower validation RMSE ($0.294 \pm 0.025$ versus $0.407 \pm 0.107$) and lower test RMSE ($0.316 \pm 0.054$ versus $0.430 \pm 0.140$), while the $\kappa_{\gamma}$-based model is substantially faster to train ($12.87 \pm 0.35$ seconds versus $83.21 \pm 1.14$ seconds). Thus, also in the full nonlinear setting, the comparison reflects the same accuracy-efficiency tradeoff observed at the core level: $\psi_{\gamma}$ provides the more accurate fit, whereas $\kappa_{\gamma}$ retains a clear computational advantage. As an external reference point, these results would place the $\psi_{\gamma}$-L2RU approximately 6th and the $\kappa_{\gamma}$-L2RU approximately 16th among the Cascaded Tanks benchmark entries available when these experiments were prepared. Together with their modest training times, this indicates that both certified architectures remain competitive within the broader nonlinear system-identification landscape.

The validation curves and representative test trajectories generated in this experiment show the same trend. The $\psi_{\gamma}$ model converges to a lower error plateau, and both models capture the dominant oscillatory behavior of the benchmark, but $\psi_{\gamma}$ tracks the sharper transients and the final decay more accurately. This is consistent with the aggregate RMSE values in Table~\ref{tab:full_ssm_tanks_best}.

Overall, the full-SSM comparison shows that the nonlinear scaffold does not erase the core-level tradeoff. Rather, it moderates it: once both parametrizations are embedded into the same architecture, $\psi_{\gamma}$ remains preferable when predictive accuracy is the main objective, while $\kappa_{\gamma}$ remains attractive when computational efficiency is the dominant concern.

\subsection{Why prescribing the $\mc L_2$-bound matters}

The previous subsections compare the two certified parametrizations proposed in this paper. We conclude by recalling why the ability to prescribe an $\mc L_2$-bound is useful in the first place. The point is not only to obtain models whose gain can be measured after training, but to make the gain bound part of the model class itself. For both $\psi_{\gamma}$ and $\kappa_{\gamma}$, every value of the free parameters produces an LTI core satisfying the prescribed bound. This turns the gain constraint into an architectural property that is preserved throughout optimization.

This is essential in the learning-based control setting of~\cite{furieri_learning_2024}. There, an L2RU controller
\[
    \Sigma_{\omega}^{\gamma}:\mc L^n \to \mc L^m
\]
is interconnected in feedback with a possibly nonlinear plant with known finite $\mc L_2$-bound $\hat{\gamma}$. The controller bound is chosen so that
\[
    \gamma \hat{\gamma} < 1,
\]
which guarantees closed-loop $\mc L_2$-stability by the small-gain theorem. Thus, $\gamma$ is a design variable, not a tuning parameter selected after the fact. Once the bound is fixed, the controller can be trained by unconstrained gradient-based optimization while the stability certificate remains valid for all iterates.

A second example is the distributed identification framework of~\cite{massai_unconstrained_2024}. In that setting, the learned model is built as an interconnection of several recurrent subsystems whose topology mirrors that of the physical plant. In the triple-tank benchmark considered there, each tank subsystem is represented by an $\mc L_2$-bounded recurrent model, and the local gain budgets are tuned so that the interconnection satisfies an overall stability condition. Here the prescribed bounds play a compositional role: they allow stability of the full learned interconnection to be certified from the individual subsystem certificates. The reported experiments show that this constraint is compatible with accurate prediction, improved long-memory training through the initialization of Proposition~\ref{prop:init}, and competitive training time compared with other $\mc L_2$-bounded recurrent architectures.

These examples highlight the limitations of unconstrained recurrent parametrizations in certified learning problems. Even if an unconstrained model achieves low prediction error on the observed trajectories, its induced gain may violate the value required by a small-gain or interconnection certificate. In contrast, a prescribed-gain parametrization enforces the relevant bound during training, independently of the sampled data, optimizer, initialization, or stopping time.

They also clarify when the two parametrizations are most useful. In closed-loop learning-based control with a nonlinear plant, the rollout is often dominated by the sequential simulation of the plant dynamics. In that case, the scan-compatible structure of $\kappa_{\gamma}$ cannot necessarily be exploited end-to-end, and the more expressive, tighter $\psi_{\gamma}$ parametrization may be preferable. In system-identification or sequence-modeling settings, instead, the learned SSM itself is often the main recurrent computation; there, the diagonal structure of $\kappa_{\gamma}$ can translate into substantial speedups, especially on long sequences.

In this sense, prescribing the $\mc L_2$-bound turns robustness from an empirical outcome into a design constraint, while $\psi_{\gamma}$ and $\kappa_{\gamma}$ provide two complementary ways of imposing that constraint. The former prioritizes expressiveness and tight use of the gain budget; the latter prioritizes diagonal structure and scan-compatible computation.

\begin{table*}[t]
\centering
\caption{Full-SSM same-scaffold comparison on the \emph{Cascaded Tanks} benchmark. For each parametrization, we report the best initialization according to the monitoring RMSE. Values are mean $\pm$ standard deviation over $4$ runs.}
\label{tab:full_ssm_tanks_best}
\small
\renewcommand{\arraystretch}{1.2}
\setlength{\tabcolsep}{14pt}
\rowcolors{2}{gray!10}{white}
\begin{tabular}{lcccc}
\toprule
\textbf{Core} &
\textbf{\# Params} &
\textbf{Test RMSE} $\downarrow$ &
\textbf{Train time [s]} $\downarrow$ &
\textbf{Learned $\gamma$} \\
\midrule
$\psi_{\gamma}$ &
$1358$ &
$\mathbf{0.316 \pm 0.054}$ &
$83.21 \pm 1.14$ &
$1.304 \pm 0.111$ \\

$\kappa_{\gamma}$ &
$1385$ &
$0.430 \pm 0.140$ &
$\mathbf{12.87 \pm 0.35}$ &
$1.904 \pm 0.326$ \\
\bottomrule
\end{tabular}
\rowcolors{2}{white}{white}
\end{table*}

\begin{figure*}[t]
\centering
\includegraphics[scale=.3]{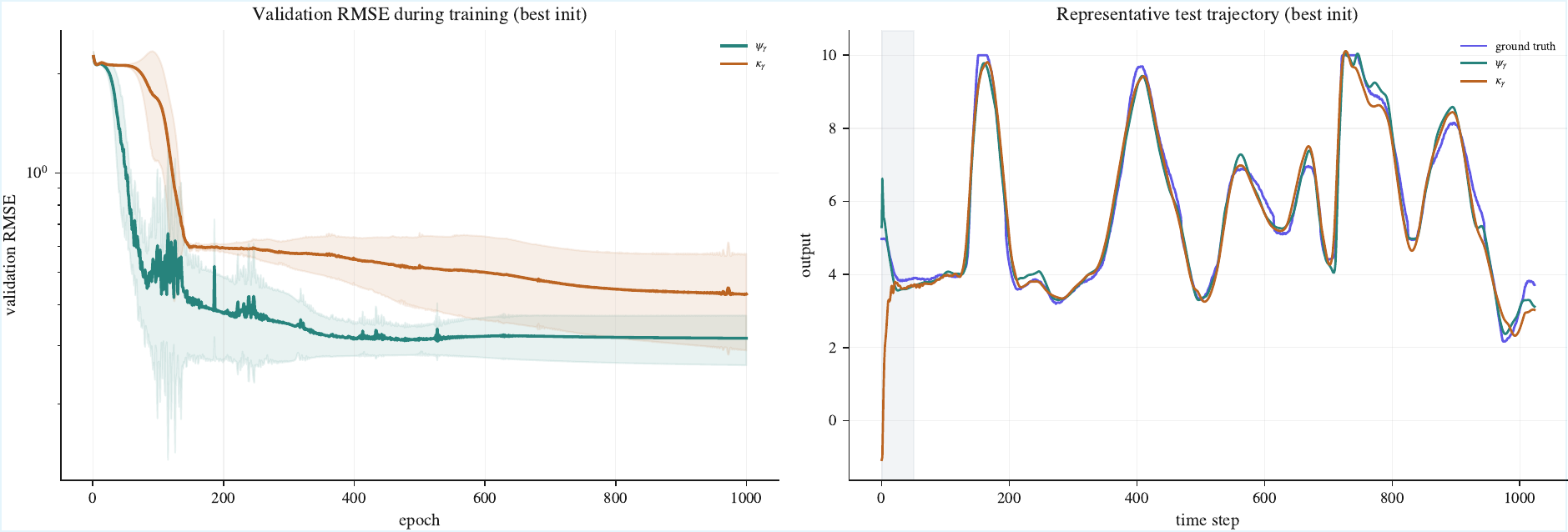}
\caption{Data-independent gain control. The prescribed-gain L2RU has a deterministic $H_\infty$ ceiling, whereas the unconstrained LTI model has training-dependent worst-case amplification.}
\label{fig:gain_control}
\end{figure*}

\section{Conclusion}

We presented L2RU, an $\mathcal{L}_2$-bounded SSM architecture designed to guarantee stability and robustness during training and deployment. Central to our approach are two free parametrizations of discrete-time LTI systems satisfying a prescribed $\mathcal{L}_2$-bound: a non-conservative parametrization offering a complete representation of square systems, and a conservative parametrization that extends the framework to general non-square systems. These tools enable L2RU to be trained through unconstrained optimization while maintaining rigorous stability and robustness guarantees, certified by the prescribed $\mc L_2$-bound. Together with a dedicated long-memory initialization strategy, they yield strong performance in system identification benchmarks.

\vspace{-2pt}

\bibliographystyle{IEEEtran}
\bibliography{biblio}

@misc{massai_free_2025,
	title = {Free {Parametrization} of {L2}-bounded {State} {Space} {Models}},
	url = {http://arxiv.org/abs/2503.23818},
	doi = {10.48550/arXiv.2503.23818},
	urldate = {2025-08-27},
	publisher = {arXiv},
	author = {Massai, Leonardo and Ferrari-Trecate, Giancarlo},
	month = mar,
	year = {2025},
	note = {arXiv:2503.23818},
}

@article{bonassi_recurrent_2022,
	title = {On {Recurrent} {Neural} {Networks} for learning-based control: {Recent} results and ideas for future developments},
	volume = {114},
	issn = {09591524},
	shorttitle = {On {Recurrent} {Neural} {Networks} for learning-based control},
	doi = {10.1016/j.jprocont.2022.04.011},
	language = {en},
	urldate = {2025-02-25},
	journal = {Journal of Process Control},
	author = {Bonassi, Fabio and Farina, Marcello and Xie, Jing and Scattolini, Riccardo},
	month = jun,
	year = {2022},
	pages = {92--104},
}

@inproceedings{sun_efficient_2022,
	address = {Bristol, United Kingdom},
	title = {Efficient {Mask} {Attention}-{Based} {NARMAX} ({MAB}-{NARMAX}) {Model} {Identification}},
	copyright = {https://doi.org/10.15223/policy-029},
	isbn = {9781665498074},
	doi = {10.1109/ICAC55051.2022.9911110},
	urldate = {2025-02-25},
	booktitle = {2022 27th {International} {Conference} on {Automation} and {Computing} ({ICAC})},
	publisher = {IEEE},
	author = {Sun, Yiming and Wei, Hua-Liang},
	month = sep,
	year = {2022},
	pages = {1--6},
}

@article{gu_efficiently_2022,
	title = {Efficiently {Modeling} {Long} {Sequences} with {Structured} {State} {Spaces}},
	doi = {10.48550/arXiv.2111.00396},
	urldate = {2025-02-25},
	publisher = {arXiv},
	author = {Gu, Albert and Goel, Karan and Ré, Christopher},
	month = aug,
	year = {2022},
	note = {arXiv:2111.00396},
}

@article{orvieto_resurrecting_2023,
	title = {Resurrecting {Recurrent} {Neural} {Networks} for {Long} {Sequences}},
	doi = {10.48550/arXiv.2303.06349},
	urldate = {2025-02-25},
	publisher = {arXiv},
	author = {Orvieto, Antonio and Smith, Samuel L. and Gu, Albert and Fernando, Anushan and Gulcehre, Caglar and Pascanu, Razvan and De, Soham},
	month = mar,
	year = {2023},
	note = {arXiv:2303.06349},
}

@article{alonso_state_2024,
	title = {State {Space} {Models} as {Foundation} {Models}: {A} {Control} {Theoretic} {Overview}},
	shorttitle = {State {Space} {Models} as {Foundation} {Models}},
	doi = {10.48550/arXiv.2403.16899},
	urldate = {2025-02-25},
	publisher = {arXiv},
	author = {Alonso, Carmen Amo and Sieber, Jerome and Zeilinger, Melanie N.},
	month = mar,
	year = {2024},
	note = {arXiv:2403.16899},
}

@article{wang_direct_2023,
	title = {Direct {Parameterization} of {Lipschitz}-{Bounded} {Deep} {Networks}},
	doi = {10.48550/arXiv.2301.11526},
	urldate = {2025-02-25},
	publisher = {arXiv},
	author = {Wang, Ruigang and Manchester, Ian R.},
	month = jun,
	year = {2023},
	note = {arXiv:2301.11526},
}

@article{revay_recurrent_2024,
	title = {Recurrent {Equilibrium} {Networks}: {Flexible} {Dynamic} {Models} {With} {Guaranteed} {Stability} and {Robustness}},
	volume = {69},
	issn = {1558-2523},
	shorttitle = {Recurrent {Equilibrium} {Networks}},
	doi = {10.1109/TAC.2023.3294101},
	number = {5},
	urldate = {2025-02-25},
	journal = {IEEE Transactions on Automatic Control},
	author = {Revay, Max and Wang, Ruigang and Manchester, Ian R.},
	month = may,
	year = {2024},
	pages = {2855--2870},
}

@article{pauli_training_2022,
	title = {Training {Robust} {Neural} {Networks} {Using} {Lipschitz} {Bounds}},
	volume = {6},
	issn = {2475-1456},
	doi = {10.1109/LCSYS.2021.3050444},
	urldate = {2025-02-25},
	journal = {IEEE Control Systems Letters},
	author = {Pauli, Patricia and Koch, Anne and Berberich, Julian and Kohler, Paul and Allgöwer, Frank},
	year = {2022},
	pages = {121--126},
}

@article{gu_mamba:_2024,
	title = {Mamba: {Linear}-{Time} {Sequence} {Modeling} with {Selective} {State} {Spaces}},
	shorttitle = {Mamba},
	doi = {10.48550/arXiv.2312.00752},
	urldate = {2025-02-25},
	publisher = {arXiv},
	author = {Gu, Albert and Dao, Tri},
	month = may,
	year = {2024},
	note = {arXiv:2312.00752},
}

@article{gu_hippo:_2020,
	title = {{HiPPO}: {Recurrent} {Memory} with {Optimal} {Polynomial} {Projections}},
	shorttitle = {{HiPPO}},
	doi = {10.48550/arXiv.2008.07669},
	urldate = {2025-02-25},
	publisher = {arXiv},
	author = {Gu, Albert and Dao, Tri and Ermon, Stefano and Rudra, Atri and Re, Christopher},
	month = oct,
	year = {2020},
	note = {arXiv:2008.07669},
}

@article{furieri_learning_2024,
	title = {Learning to {Boost} the {Performance} of {Stable} {Nonlinear} {Systems}},
	volume = {3},
	issn = {2694-085X},
	doi = {10.1109/OJCSYS.2024.3441768},
	urldate = {2025-02-25},
	journal = {IEEE Open Journal of Control Systems},
	author = {Furieri, Luca and Galimberti, Clara Lucía and Ferrari-Trecate, Giancarlo},
	year = {2024},
	pages = {342--357},
}

@inproceedings{furieri_distributed_2022,
	title = {Distributed {Neural} {Network} {Control} with {Dependability} {Guarantees}: a {Compositional} {Port}-{Hamiltonian} {Approach}},
	shorttitle = {Distributed {Neural} {Network} {Control} with {Dependability} {Guarantees}},
	language = {en},
	urldate = {2025-02-25},
	booktitle = {Proceedings of {The} 4th {Annual} {Learning} for {Dynamics} and {Control} {Conference}},
	publisher = {PMLR},
	author = {Furieri, Luca and Galimberti, Clara Lucía and Zakwan, Muhammad and Ferrari-Trecate, Giancarlo},
	month = may,
	year = {2022},
	pages = {571--583},
}

@inproceedings{massai_unconstrained_2024,
	title = {Unconstrained {Learning} of {Networked} {Nonlinear} {Systems} via {Free} {Parametrization} of {Stable} {Interconnected} {Operators}},
	doi = {10.23919/ECC64448.2024.10591242},
	urldate = {2025-02-25},
	booktitle = {2024 {European} {Control} {Conference} ({ECC})},
	author = {Massai, Leonardo and Saccani, Danilo and Furieri, Luca and Ferrari-Trecate, Giancarlo},
	month = jun,
	year = {2024},
	pages = {651--656},
}

@article{saccani_optimal_2024,
	title = {Optimal distributed control with stability guarantees by training a network of neural closed-loop maps},
	doi = {10.48550/arXiv.2404.02820},
	urldate = {2025-02-25},
	publisher = {arXiv},
	author = {Saccani, Danilo and Massai, Leonardo and Furieri, Luca and Ferrari-Trecate, Giancarlo},
	month = jul,
	year = {2024},
	note = {arXiv:2404.02820},
}

@inproceedings{blelloch_prefix_2004,
	title = {Prefix sums and their applications},
	copyright = {In Copyright},
	doi = {10.1184/R1/6608579.V1},
	urldate = {2025-02-25},
	publisher = {Carnegie Mellon University},
	author = {Blelloch, Guy E.},
	year = {2004},
	pages = {1294199 Bytes},
}

@article{gu_parameterization_2022,
	title = {On the {Parameterization} and {Initialization} of {Diagonal} {State} {Space} {Models}},
	doi = {10.48550/arXiv.2206.11893},
	urldate = {2025-02-25},
	publisher = {arXiv},
	author = {Gu, Albert and Gupta, Ankit and Goel, Karan and Ré, Christopher},
	month = aug,
	year = {2022},
	note = {arXiv:2206.11893},
}

@misc{caverly_lmi_2024,
	title = {{LMI} {Properties} and {Applications} in {Systems}, {Stability}, and {Control} {Theory}},
	doi = {10.48550/arXiv.1903.08599},
	urldate = {2025-02-25},
	publisher = {arXiv},
	author = {Caverly, Ryan James and Forbes, James Richard},
	month = may,
	year = {2024},
	note = {arXiv:1903.08599},
}

@misc{smith_simplified_2023,
	title = {Simplified {State} {Space} {Layers} for {Sequence} {Modeling}},
	doi = {10.48550/arXiv.2208.04933},
	urldate = {2025-02-25},
	publisher = {arXiv},
	author = {Smith, Jimmy T. H. and Warrington, Andrew and Linderman, Scott W.},
	month = mar,
	year = {2023},
	note = {arXiv:2208.04933},
}

@misc{vaswani_attention_2023,
	title = {Attention {Is} {All} {You} {Need}},
	doi = {10.48550/arXiv.1706.03762},
	urldate = {2025-02-25},
	publisher = {arXiv},
	author = {Vaswani, Ashish and Shazeer, Noam and Parmar, Niki and Uszkoreit, Jakob and Jones, Llion and Gomez, Aidan N. and Kaiser, Lukasz and Polosukhin, Illia},
	month = aug,
	year = {2023},
	note = {arXiv:1706.03762},
}

@inproceedings{andersson_deep_2019,
	title = {Deep {Convolutional} {Networks} in {System} {Identification}},
	doi = {10.1109/CDC40024.2019.9030219},
	urldate = {2025-02-25},
	booktitle = {2019 {IEEE} 58th {Conference} on {Decision} and {Control} ({CDC})},
	author = {Andersson, Carl and Ribeiro, Antônio H. and Tiels, Koen and Wahlström, Niklas and Schön, Thomas B.},
	month = dec,
	year = {2019},
	note = {ISSN: 2576-2370},
	pages = {3670--3676},
}

@inproceedings{lanzetti_recurrent_2019,
	title = {Recurrent {Neural} {Network} based {MPC} for {Process} {Industries}},
	doi = {10.23919/ECC.2019.8795809},
	urldate = {2025-02-25},
	booktitle = {2019 18th {European} {Control} {Conference} ({ECC})},
	author = {Lanzetti, Nicolas and Lian, Ying Zhao and Cortinovis, Andrea and Dominguez, Luis and Mercangöz, Mehmet and Jones, Colin},
	month = jun,
	year = {2019},
	pages = {1005--1010},
}

@article{bonassi_structured_2024,
	series = {20th {IFAC} {Symposium} on {System} {Identification} {SYSID} 2024},
	title = {Structured state-space models are deep {Wiener} models},
	volume = {58},
	issn = {2405-8963},
	doi = {10.1016/j.ifacol.2024.08.536},
	number = {15},
	urldate = {2025-02-25},
	journal = {IFAC-PapersOnLine},
	author = {Bonassi, Fabio and Andersson, Carl and Mattsson, Per and Schön, Thomas B.},
	month = jan,
	year = {2024},
	pages = {247--252},
}

@article{zakwan_robust_2023,
	title = {Robust {Classification} {Using} {Contractive} {Hamiltonian} {Neural} {ODEs}},
	volume = {7},
	issn = {2475-1456},
	doi = {10.1109/LCSYS.2022.3186959},
	urldate = {2025-03-11},
	journal = {IEEE Control Systems Letters},
	author = {Zakwan, Muhammad and Xu, Liang and Ferrari-Trecate, Giancarlo},
	year = {2023},
	pages = {145--150},
}

\end{document}